\documentclass[prd,aps,preprint,tightenlines,superscriptaddress]{revtex4}
\usepackage{amssymb}
\usepackage{epsfig}
\usepackage{amsmath}
\begin{document}
\title{Right-Handed Quark Mixings in Minimal \\ Left-Right Symmetric
Model with General CP Violation}
\author{Yue Zhang}
\affiliation{Center for High-Energy Physics and Institute of
Theoretical Physics, Peking University, Beijing 100871, China}
 \affiliation{Department of Physics, University of
Maryland, College Park, Maryland 20742, USA }
\author{Haipeng An}
 \affiliation{Department of Physics, University of
Maryland, College Park, Maryland 20742, USA }
\author{Xiangdong Ji}
 \affiliation{Department of Physics, University of
Maryland, College Park, Maryland 20742, USA } \affiliation{Center
for High-Energy Physics and Institute of Theoretical Physics, Peking
University, Beijing 100871, China}
\author{R. N. Mohapatra}
 \affiliation{Department of Physics, University of
Maryland, College Park, Maryland 20742, USA }
\date{\today}
\begin{abstract}

We present a systematic approach to solve analytically for the right-handed quark mixings
in the minimal left-right symmetric model which generally has both explicit and
spontaneous CP violations. The leading-order result has the same hierarchical structure
as the left-handed CKM mixing, but with additional CP phases originating from a
spontaneous CP-violating phase in the Higgs vev. We explore the phenomenology entailed by
the new right-handed mixing matrix, particularly the bounds on the mass of $W_R$ and the
CP phase of the Higgs vev.

\end{abstract}
\maketitle

The physics beyond the standard model (SM) has been the central focus of high-energy
phenomenology for more than three decades. Many proposals, including supersymmetry,
technicolors, little Higgs, and extra dimensions, have been made and studied thoroughly
in the literature; tests are soon to be made at the Large Hadron Collider (LHC). One of
the earliest proposals, the left-right symmetric (LR) model \cite{lrmodel}, was motivated
by the hypothesis that parity is a perfect symmetry at high-energy, and is broken
spontaneously at low-energy due to an asymmetric vacuum. Asymptotic restoration of parity
has a definite aesthetic appeal \cite{lee}. This model, based on the gauge group
$SU(2)_L\times SU(2)_R\times U(1)_{B-L}$, has a number of additional attractive features,
including a natural explanation of the weak hyper-change in terms of baryon and lepton
numbers \cite{marshak}, the existence of right-handed neutrinos, and the possibility of
spontaneous CP (charge-conjugation-parity) violation (SCPV) \cite{lee1}. The model can
easily be constrained by low-energy physics and predict clear signatures at colliders
\cite{cmsnote}. It so far remains a decent possibility for new physics.

The LR modes are best constrained at low-energies by flavor-changing mixings and decays,
particularly the CP violating observables. In making theoretical predictions, the major
uncertainty comes from the unknown right-handed quark mixing matrix, conceptually similar
to the left-handed quark Cabibbo-Kobayashi-Maskawa (CKM) mixing. The new mixing generally
depends on 9 real parameters: 6 CP violation phases and 3 rotational angles. Over the
years, two limiting cases of the model have usually been studied. The first case,
``manifest left-right symmetry", assumes that there is no SCPV, i.e., all Higgs vacuum
expectation values (vev) are real. The quark mass matrices are then hermitian, and the
left and right-handed quark mixings become identical, modulo the sign uncertainty from
possible negative quark masses. The reality of the Higgs vev, however, does not survive
radiative corrections which generate infinite renormalization. The second case,
``pseudo-manifest left-right symmetry", assumes that the CP violation comes entirely from
spontaneous symmetry breaking (SSB) and that all Yukawa couplings are real \cite{scpv}.
Here the quark mass matrices are complex but symmetric, the right-handed quark mixing is
related to the complex conjugate of the CKM matrix multiplied by additional CP phases.
There are few studies of the model with general CP violation in the literature
\cite{generalcp}, with the exception of an extensive numerical study in Ref. \cite{kiers}
where solutions were generated through a Monte Carlo method.

In this paper, we report a systematic approach to solve analytically for the right-handed
quark mixings in the minimal LR model with general CP violation. As is well-known, the
model has a Higgs bi-doublet whose vev's are complex, leading to both explicit and
spontaneous CP violations. The approach is based on the fact that $m_t\gg m_b$ and hence
the ratio of the two vev's of the Higgs bi-doublet, $\xi = \kappa'/\kappa$, is small. In
the leading-order in $\xi$, we find a linear matrix equation for the right-handed quark
mixing which can readily be solved. We present an analytical solution of this equation
valid to ${\cal O}(\lambda^3)$, where $\lambda=\sin\theta_C$ is the Cabibbo mixing
parameter. The leading-order solution is very close to the left-handed CKM matrix, apart
from additional phases that are fixed by $\xi$, the spontaneous CP phase $\alpha$, and
the quark masses. This explicit right-handed quark mixing allows definitive studies of
the neutral meson mixing and CP-violating observables. We use the experimental data on
kaon and $B$-meson mixings and neutron electrical dipole moment (EMD) to constrain the
mass of $W_R$ and the SCPV phase $\alpha$.

The matter content of the LR model is the same as the standard model (SM), except for a
right-handed neutrino for each family which, together with the right-handed charged
lepton, forms a $SU(2)_R$ doublet. The Higgs sector contains a bi-doublet $\phi$, which
transforms like (2,2,0) of the gauge group, and the left and right triplets
$\Delta_{L,R}$, which transform as $(3,1,2)$ and $(1,3,2)$, respectively. The gauge group
is broken spontaneously into the SM group $SU(2)_L\times U(1)_Y$ at scale $v_R$ through
the vev of $\Delta_R$. The breaking of the SM group is accomplished through vev's of
$\phi$.

The most general renormalizable Higgs potential can be found in Ref. \cite{pot}. Only one
of the parameters, $\alpha_2$, which describes an interaction between the bi-doublet and
triplet Higgs, is complex, and induces an explicit CP violation in the Higgs potential.
It is known in the literature that when this parameter is real, SCPV does not occur if
the SM group is to be recovered in the decoupling limit $v_R\rightarrow\infty$
\cite{pot}. Without SCPV, the Yukawa couplings in the quark sector are hermitian, and we
have the manifest left-right symmetry limit. Here we are interested in the general case
when $\alpha_2$ is complex. A complex $\alpha_2$ allows spontaneous CP violation as well,
generating a finite phase $\alpha$ for the vevs of $\phi$,
\begin{equation}
  \langle \phi\rangle =\left( \begin{array}{cc}
                \kappa   &  0 \\
                 0    &  \kappa'e^{i\alpha} \end{array}\right) \ .
\end{equation}
In reference \cite{pot1}, a relation was derived between $\alpha$
and the phase $\delta_2$ of $\alpha_2$,
\begin{equation}
    \alpha \sim
    \sin^{-1}\left(\frac{2|\alpha_2|\sin\delta_2}{\alpha_3}\xi\right)  \ ,
\end{equation}
where $\alpha_3$ is another interaction parameter between the Higgs bi-doublet and
triplets.

The quark masses in the model are generated from the Yukawa coupling,
\begin{equation}
  {\cal L}_Y =  \bar q (h \phi + \tilde h \tilde \phi) q + {\rm h.
  c.} \ .
\end{equation}
Parity symmetry ($\phi\rightarrow \phi^\dagger$, $q_L\rightarrow q_R$) constrains $h$ and
$\tilde h$ be hermitian matrices. After SSB, the above lagrangian yields the following
quark mass matrices,
\begin{eqnarray}
     M_u &=& \kappa h + \kappa' e^{-i\alpha} \tilde h \nonumber \\
     M_d &=& \kappa' e^{i\alpha} h  + \kappa \tilde h\ .
\end{eqnarray}
Because of the non-zero $\alpha$, both $M_u$ and $M_d$ are non-hermitian. And therefore,
the right-handed quark mixing can in principle very different from that of the left-hand
counter part.

Since the top quark mass is much larger than that of down quark, one may assume, without
loss of generality, $\kappa'\ll\kappa$, while at the same time $\tilde h$ is at most the
same order as $h$. We parameterize $\kappa'/\kappa = r m_b/m_t$, where $r$ is a parameter
of order unity. As a consequence, $M_u$ is nearly hermitian, and one may neglect the
second term to leading order in $\xi$. One can account for it systematically in $\xi$
expansion if the precision of a calculation demands. Now $h$ can be diagonalized by a
unitary matrix $U_u$,
\begin{equation}
   M_u = U_u \hat M_u SU_u^\dagger = \kappa h \ ,
\end{equation}
where $\hat M_u$ is diag$(m_u,m_c,m_t)$, and $S$ is a diagonal sign
matrix, diag$(s_u,s_c,s_t)$, satisfying $S^2=1$. Replacing the
$h$-matrix in $M_d$ by the above expression, one finds
\begin{equation}
    e^{i\alpha} \xi \hat M_u
  + \kappa U_u\tilde h U^\dagger_uS
     = V_L \hat M_d V_R^\dagger
\end{equation}
where $\hat M_d$ is diag$(m_d,m_s,m_b)$, $V_L$ is the CKM matrix and $V_R$ is the
right-handed mixing matrix that we are after. Two comments are in order. First, through
redefinitions of quark fields, one can bring $V_L$ to the standard CKM form with four
parameters (3 rotations and 1 CP violating phase) and the above equation remains the
same. Second, all parameters in the unitary matrix $V_R$ are now physical, including 3
rotations and 6 CP-violating phases.

To make further progress, one uses the hermiticity condition for
$U_u \tilde h U_u^\dagger$, which yields the following equation,
\begin{equation}
  \hat M_d\hat V_R^\dagger - \hat V_R \hat
    M_d = 2i\xi\sin\alpha ~ V^\dagger_L \hat M_u S V_L
\end{equation}
where $\hat V_R$ is the quotient between the left and right mixing $V_R = S V_L\hat V_R$.
There are a total of 9 equations above, which are sufficient to solve 9 parameters in
$\hat V_R$. It is interesting to note that if there is no SCPV, $\alpha=0$, the solution
is simply $V_R = S V_L \tilde S$, where $\tilde S$ is another diagonal sign matrix,
diag$( s_d,s_s,s_b)$,  satisfying $\tilde S^2 =1$. We recover the manifest left-right
symmetry case.

The above linear equation can be solved using various methods. The simplest is to utilize
the hierarchy between down-type-quark masses. Multiplying out the left-hand side and
assuming $\hat V_{Rij}$ and $\hat V_{Rij}^*$ are of the same order, which can be
justified posteriori, the solution is (for $\rho\sin\alpha\le 1$)
\begin{eqnarray}
  {\rm Im}\hat V_{R11} &=& -r\sin\alpha\frac{m_bm_c}{m_dm_t}\lambda^2
  \nonumber \\
    &\times& \left(s_c + s_t\frac{m_t}{m_c}A^2\lambda^4((1-\rho)^2+\eta^2)\right) \\
  {\rm Im}\hat V_{R22} &=&  - r\sin\alpha \frac{m_bm_c}{m_sm_t}
    \left(s_c + s_t\frac{m_t}{m_c}A^2\lambda^4 \right)\\
  {\rm Im} \hat V_{R33} &=& -r\sin\alpha~ s_t \\
 \hat  V_{R12} &=& 2i r \sin\alpha \frac{m_bm_c}{m_sm_t} \lambda
   \nonumber \\
   &\times& \left(s_c + s_t\frac{m_t}{m_c}\lambda^4A^2(1-\rho+i\eta)\right)  \\
  \hat V_{R13} &=& -2i r \sin\alpha A\lambda^3(1-\rho+i\eta)s_t\\
  \hat V_{R23} &=& 2i r \sin\alpha A\lambda^2 s_t \ ,
\end{eqnarray}
where Im$\hat V_{R11}$, Im$\hat V_{R22}$, Im$\hat V_{R33}$, $\hat V_{R12}$, $\hat
V_{R23}$, $\hat V_{R13}$ are on the orders of $\lambda$, $\lambda$, $1$, $\lambda^2$,
$\lambda^2$, and $\lambda^3$, respectively. The above solution allows us to construct
entirely the right-handed mixing to order ${\cal O}(\lambda^3)$
\begin{equation}
     V_R = P_U V P_D  \ ,
\end{equation}
where the factors $P_U=$diag$(s_u, s_d\exp(2i\theta_2),s_t\exp(2i\theta_3))$,
$P_D=$diag$(s_d\exp(i\theta_1), s_s\exp(-i\theta_2),s_b\exp(-i\theta_3))$, and
\begin{eqnarray}
 V =\left(\begin{array}{ccc}
               1-\lambda^2/2 & \lambda & A\lambda^3(\rho-i\eta) \\
                 -\lambda & 1-\lambda^2/2 & A\lambda^2e^{-i2\theta_2}  \\
                 A\lambda^3(1-\rho-i\eta) & -A\lambda^2e^{2i\theta_2} & 1 \end{array}
             \right);
\end{eqnarray}
with $\theta_i =\tilde s_i \sin^{-1} {\rm Im}\hat V_{Rii}$. The pseudo-manifest limit is
recovered when $\eta=0$.

A few remarks about the above result are in order. First, the hierarchical structure of
the mixing is similar to that of the CKM, namely 1-2 mixing is of order $\lambda$, 1-3
order $\lambda^3$ and 2-3 order $\lambda^2$. Second, every element has a significant CP
phase. The elements involving the first two families have CP phases of order $\lambda$,
and the phases involving the third family are of order $1$. These phases are all related
to the single SCPV phase $\alpha$, and can produce rich phenomenology for $K$ and $B$
meson systems as well as the neutron EDM. Third, depending on signs of the quark masses,
there are $2^5=32$ discrete solutions. Finally, using the right-handed mixing at leading
order in $\xi$, one can construct $\tilde h$ from Eq. (6) and solve $M_u$ with a better
approximation. The iteration yields a systematic expansion in $\xi$.

In the remainder of this paper, we consider the kaon and $B$-meson mixing as well as the
neutron EMD. We will first study the contribution to the $K_L-K_S$ mass difference
$\Delta M_K$ and derive an improved bound on the mass of right-handed gauge boson $W_R$,
using the updated hadronic matrix elements and strange quark mass. Then we calculate the
CP violation parameter $\epsilon$ in $K_L$ decay and the neutron EDM, deriving an
independent bound on $M_{W_R}$. Finally, we consider the implications of the model in the
$B$-meson system, deriving yet another bound on $M_{W_R}$.

The leading non-SM contribution to the $K_0-\overline{K}_0$ mixing comes from the
$W_L-W_R$ box diagram and the tree-level flavor-changing, neutral-Higgs (FCNH)
diagram\cite{bounds, kaon2}. The latter contribution has the same sign as the former, and
inversely proportional to the square of the FCNH boson masses. We assume large Higgs
boson masses $(> 20$TeV) from a large $\alpha_3$ in the Higgs potential so that the
contribution to the mixing is negligible. Henceforth we concentrate on the box diagram
only.

Because of the strong hierarchical structure in the left and right
quark mixing, the $W_L-W_R$ box contribution to the kaon mixing
comes mostly from the intermediate charm quark,
\begin{eqnarray}
   H_{12} &=& \frac{G_F}{2}
    \frac{\alpha_{\rm em}}{4\pi\sin^2\theta_W} 2\eta \lambda_c^{LR}
    \lambda^{RL}_c m_c^2
    \\ &\times& [4(1+\ln x_c)+ \ln\eta]
      ~ \left[(\bar ds)^2-(\bar d\gamma_5s)^2\right]
      + {\rm  h. c.} \nonumber
\end{eqnarray}
where $x_c = m_c^2/M_{W_L}^2$, $\eta = M_{W_L}^2/M_{W_R}^2$, $\lambda^{RL}_c=
V_{Rcd}^*V_{Lcs}$, and $\lambda^{LR}_c = V_{Lcd}^*V_{Rcs}$. The above result is very
similar to that from the manifest-symmetry limit because the phases in $V_{Rcd}$ and
$V_{Rcs}$ are ${\cal O}(\lambda)$. Therefore, we expect a similar bound on $M_{W_R}$ as
derived in previous work \cite{bounds}. However, the rapid progress in lattice QCD
calculations warrants an update. When the QCD radiative corrections are taken into
account explicitly, the above effective hamiltonian will be multiplied by an additional
factor $\eta_4$. [We neglect contributions of other operators with small coefficients.]
In the leading-logarithmic approximation, $\eta_4$ is about 1.4 when the the four-quark
operators are defined at the scale of 2 GeV in $\overline{\rm MS}$ scheme \cite{running}.

The hadronic matrix element of the above operator can be calculated
in lattice QCD and expressed in terms of a factorized form
\begin{eqnarray}
&& \langle K_0|\bar d(1-\gamma_5)s\bar d(1+\gamma_5)s
|\overline{K}_0\rangle \nonumber \\
&=& 2M_K f_K^2B_4(\mu) \left(\frac{m_K}{m_s(\mu)+m_d(\mu)}\right)^2 \ .
\end{eqnarray}
Using the domain-wall fermion, one finds $B_4 = 0.81$ at $\mu=2$ GeV in naive dimensional
regularization (NDR) scheme \cite{Bfactor}. In the same scheme and scale, the strange
quark mass is $m_s = 98(6)$ MeV. Using the standard assumption that the new physics
contribution shall be less than the experimental value, one finds
\begin{equation}
    M_{W_R} > 2.5 ~{\rm TeV} \ ,
\end{equation}
which is now {\it the} bound in the model with the general CP violation. This bound is
stronger than similar ones obtained before because of the new chiral-symmetric
calculation of $B_4$ and the updated value of the strange quark mass.

The most interesting predictions of $V_R$ are for CP-violating observables. We first
study the CP violating parameter $\epsilon$ in $K_L$ decay. When the SCPV phase
$\alpha=0$, the $W_L-W_R$ box diagram still makes a significant contribution to
$\epsilon$ from the phase $\delta_{\rm CKM}$ of the CKM matrix. The experimental data
then requires $W_R$ be at least 20 TeV to suppress this contribution. When $\alpha\ne 0$,
it is possible to relax the constraint by cancelations. The most significant contribution
due to $\alpha$ comes from the element $V_{Rcd}$ which is naturally on the order of
$\lambda$. In the presence of $\alpha$, we have an approximate expression for
$\epsilon_{LR}$
\begin{equation}
 \epsilon_{LR} = 0.77 \left( \frac{1~{\rm TeV}}{M_R} \right)^2 s_s  s_d ~{\rm Im}
 \left[ g(M_R, \theta_2, \theta_3) e^{-i(\theta_1 + \theta_2)} \right]
\end{equation}
where the function $g(M_R, \theta_2, \theta_3) = -2.22 + [ 0.076 + (0.030 + 0.013 i) \cos
2 (\theta_2 - \theta_3) ] \ln \left( \frac{80~{\rm GeV}}{M_R} \right)^2$. The required
value of $r\sin\alpha$ for cancelation depends sensitively on the sign of quark masses.
When $s_s=s_d$, there exist small $r \sin \alpha$ solutions even when $M_{W_R}$ is as low
as 1 TeV, as shown by solid dots in Fig. 1. However, when $s_s=-s_d$, only large $r \sin
\alpha$ solutions are possible.


\begin{figure}[hbt]
\begin{center}
\includegraphics[width=8cm]{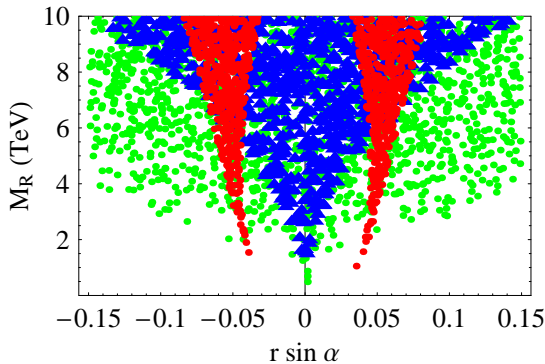}
\end{center}
\caption{Constaints on the mass of $W_R$ and the spontaneous CP violating parameter
$r\sin\alpha$ from $\epsilon$ (red dots) and neutron EDM in two different limit: small
$r\sim 0.1$ (green dots) and large $r\sim 1$ (blue trinagles).} \label{Feynman-dg2}
\end{figure}
An intriguing feature appears if one considers the constraint from the neutron EMD as
well. A calculation of EDM is generally complicated because of strongly interacting
quarks inside the neutron. As an estimate, one can work in the quark models by first
calculating the EDM of the constituent quarks. In our model, there is a dominant
contribution from the $W_L-W_R$ boson mixing \cite{edm}. Requiring the theoretical value
be below the current experimental bound, the neutron EDM prefers small $r \sin \alpha$
solutions as can be seen in Fig. 1. The combined constraints from $\epsilon$ and the
neutron EMD $(d_n <3\times 10^{-26}$ {\rm e~cm} \cite{edmbound}) impose an independent
bound on $M_{W_R}$
\begin{equation}
    M_{W_R} > (2-6) ~{\rm TeV} \ ,
\end{equation}
where the lowest bound is obtained for small $r\sim 0.05$ and large CP phase
$\alpha=\pi/2$.


Finally we consider the neutral $B$-meson mixing and CP-violating decays. In
$B_d-\overline{B}_d$ and $B_s-\overline{B}_s$ mixing, due to the heavy b-quark mass,
there is no chiral enhancement in the hadronic matrix elements of $\Delta B=2$ operators
from $W_L-W_R$ box diagram as in the kaon case. One generally expects the constraint from
neutral $B$-meson mass difference to be weak. In fact, we find a lower bound on $W_R$
mass of 1-2 TeV from $B$-mixing. On the other hand, CP asymmetry in decay $B_d
\rightarrow J/\psi K_S$, $S_{J/\psi K_S}=\sin 2 \beta$ in the standard model, receives a
new contribution from both $B_d-\overline{B}_d$ and $K_0 -\overline{K}_0$ mixing in the
presence of $W_R$ \cite{generalcp} and is very sensitive to the relative sign $s_d$ and
$s_s$. By demanding the modified $\sin 2 \beta_{\rm eff}$ within the experimental error
bar, we find another independent bound on $M_{W_R}$,
\begin{equation}
    M_{W_R} > 2.4 ~{\rm TeV} \ ,
\end{equation}
when $s_d=s_s$ as required by the neutron EDM bound.

To summarize, we have derived analytically the right-handed quark mixing in the minimal
left-right symmetric model with general CP violation. Using this and the kaon and B-meson
mixing and the neutron EMD bound, we derive new bounds on the mass of right-handed gauge
boson, consistently above 2.5 TeV. To relax this constraint, one can consider models with
different Higgs structure and/or supersymetrize the theory. A more detailed account of
the present work, including direct CP observables, will be published elsewhere
\cite{zhang}

This work was partially supported by the U. S. Department of Energy via grant
DE-FG02-93ER-40762. Y. Z. acknowledges the hospitality and support from the TQHN group at
University of Maryland and a partial support from NSFC grants 10421503 and 10625521. X.
J. is supported partially by a grant from NSFC and a ChangJiang Scholarship at Peking
University. R. N. M. is supported by NSF grant No. PHY-0354407.

\end{document}